\documentclass[twocolumn,epjc3]{svjour3}

\usepackage{amsmath}
\usepackage{euscript,amssymb,epsfig}
\usepackage{amsfonts,latexsym}
\usepackage{color}

\newcommand{\be}[1]{\begin{equation}\label{#1}}
\newcommand{\ee}{\end{equation}}
\newcommand{\ba}[1]{\begin{eqnarray}\label{#1}}
\newcommand{\ea}{\end{eqnarray}}
\newcommand{\rf}[1]{(\ref{#1})}
\newcommand{\nn}{\nonumber}

\newcommand{\de}{\partial}

\newcommand{\Om}{\Omega}

\newcommand{\hg}{\hat g}
\newcommand{\hna}{\hat \nabla}
\newcommand{\bmu}{\bar\mu}
\newcommand{\bnu}{\bar\nu}

\journalname{Eur. Phys. J. C}

\begin{document}

\title{Weak-field limit of Kaluza-Klein models with
spherically symmetric static scalar field: observational constraints}

\author{Alexander Zhuk\thanksref{addr1,addr2} \and Alexey Chopovsky\thanksref{addr2} \and Seyed Hossein Fakhr\thanksref{addr2} \and Valerii Shulga\thanksref{addr1,addr3} \and Han Wei\thanksref{addr1}}







\institute{The International Center of Future Science of the Jilin University, 2699 Qianjin St., 130012, Changchun City, China \label{addr1}
\and
Astronomical Observatory, Odessa National University, Dvoryanskaya st. 2, Odessa 65082, Ukraine \label{addr2}
\and
Institut of Radio Astronomy of National Academy of Sciences of Ukraine, 4 Mystetstv str.,61002 Kharkov, Ukraine \label{addr3}}

\date{Received: date / Accepted: date}

\maketitle

\begin{abstract}
In a multidimensional Kaluza-Klein model with Ricci-flat internal space, we study the gravitational field in the weak-field limit. This field is created by two
coupled sources. First, this is a point-like massive body which has a dust-like equation of state in the external space and an arbitrary parameter $\Omega$ of
equation of state in the internal space.  The second source is a static spherically symmetric massive scalar field centered at the origin where the point-like
massive body is. The found perturbed metric coefficients are used to calculate the parameterized post-Newtonian (PPN) parameter $\gamma$. We define under which
conditions $\gamma$ can be very close to unity in accordance with the relativistic gravitational tests in the Solar system. This can take place for both massive
or massless scalar fields. For example, to have $\gamma \approx 1$ in the Solar system, the mass of scalar field should be $\mu \gtrsim 5.05\times 10^{-49}$g
$\sim 2.83\times 10^{-16}$eV. In all cases, we arrive at the same conclusion that to be in agreement with the relativistic gravitational tests, the gravitating
mass should have tension: $\Omega=-1/2$.
\end{abstract}

\maketitle

\vspace{.5cm}



\section{Introduction}

\setcounter{equation}{0}

Any physical theory is viable only when it does not contradict the experimental data. Theories of gravity are not an exception to this rule. It is well known that
General Relativity (GR) successfully passed the test by the gravitational experiments, e.g. the relativistic gravitational tests in the Solar system: the
deflection of light, the time delay of radar echoes and the perihelion precession of Mercury. Consequently, any modified theories of gravity must satisfy these
observations with an accuracy not less than GR. Multidimensional Kaluza-Klein (KK) models are among such modified theories of gravity.

Therefore, in our previous papers \cite{EZ1,CEZ1,CEZ2}, we investigated this problem for multidimensional KK models with compact Ricci-flat (or toroidal, as a
particular example) internal spaces. We considered the post-Newtonian gravitational field generated by discrete massive sources with negligible (in comparison to
$c$) velocities. We assumed that massive bodies possess zero pressure in the external (our) space. Here, the pressure is understood as a characteristic of a
gravitating body. It is well known that the pressure inside the non-relativistic astrophysical bodies (such as our Sun) is much less than the energy density.
Therefore, we can neglect it. This is a natural assumption in General Relativity \cite{Landau} accepted for calculation of the parameterized post-Newtonian (PPN)
parameters \cite{Will}. It is natural to assume that in the internal space the gravitating mass also has a dust-like equation of state. However, since we do not
know the equation of state in the internal space, we assumed for generality some nonzero parameter $\Omega$ in this equation. Then, for such KK model we
investigated the PPN parameter $\gamma$. This parameter is well defined from the Shapiro time-delay experiment: $\gamma=1+(2.1\pm2.3)\times 10^{-5}$
\cite{Bertotti,Shapiro,Will2}. To our surprise, in the case of KK models with Ricci-flat internal spaces, we found that to get $\gamma$ close to unity, the
equation of state parameter $\Omega$ should be very close to $-1/2$. Strictly speaking, to have $\gamma=1$ (as in General Relativity), $\Omega$ should be exactly
equal to $-1/2$. This value is well known for black strings and black branes \cite{TF,TZ,HO,TK,EZ2,Braz}. For KK models with Ricci-flat internal spaces, this result does not
depend on the size of the extra dimensions \cite{EZ12b}.{\footnote{It is a common knowledge that KK theory reduces to general relativity if the compact radii go to zero. Really, the compactness and smallness of the internal spaces is a necessary condition for KK models to be in agreement with the observations. However, this is not a sufficient condition. Since the variations of the volume of internal spaces contribute to the perturbations of the metric, then this results in the fifth force/radions. Such fifth force contradicts the observations. In the case of background Ricci-flat internal spaces, radions are massless, and this property does not depend on the size of the internal space. If the gravitating masses are the only matter sources perturbing the background geometry, to eliminate such massless radions (i.e. to eliminate the contributions of the fluctuations of the volume of internal spaces to the metric perturbations) the gravitating masses should have tension $\Omega=-1/2$ in the internal space. This is the black branes/strings case.  These models satisfy the gravitational tests in the Solar system at the same level of accuracy as general relativity. In the case of curved internal spaces we have a different picture. Here, we can also eliminate radions if $\Omega=-1/2$. However, for such models radions are massive with masses inversely proportional to the volume of the internal space, and the contributions of the fifth force have a Yukawa type form. Therefore, for an arbitrary value of $\Omega$ such KK theory reduces to general relativity in the limit of large masses (i.e. small internal space volume) where the Yukawa contributions are exponentially small. These aspects of KK models are discussed in detail in our papers \cite{EZ12b,CEZ3}.}}

{Since up to now there is no a physically motivated explanation of the origin of such essentially non-zero value of the parameter $\Omega$,} we are asking about the possibility of constructing the KK models, for which, on the one
hand, the PPN parameter $\gamma$ will satisfy the experimental constraints (i.e. will be very close to 1), and, on the other hand, the parameter $\Omega$ will be
0. This is the main aim of the present paper. To achieve this goal, we include into consideration a static spherically symmetric massive scalar field coupled with
the gravitating body. Unfortunately, we demonstrate that to have the parameter $\gamma \approx 1$ in the considered model, we should still demand $\Omega =-1/2$.

The paper is structured as follows. In Sec. 2, we describe the model and present basic equations. In Sec. 3, we calculate the perturbed metric coefficients. This
coefficients enable us to estimate the PPN parameter $\gamma$ in Sec. 4.  The main results are briefly summarized in the concluding Sec. 5.

\section{The model and basic equations}

\setcounter{equation}{0}

The background spacetime is supposed to have a $(1+D)=(1+3+d)$-dimensional block-diagonal metrics of the form
\be{1}
\hat g=\hat g_{MN}dx^M\otimes dx^N=\eta_{\mu\nu}dx^\mu\otimes dx^\nu+\hat g_{mn}dx^m\otimes dx^n\, ,
\ee
where $\eta_{\mu\nu}=\text{diag}(+1, -1, -1, -1)$ is Minkowski metric of the visible (external) 4-dimensional spacetime, and $\hat g_{mn}$ is Ricci-flat metric of
the compact $d$-dimensional extra (internal) subspace. Hereafter, capital Latin indices run from $0$ to $D$, Greek indices run from $0$ to $3$, small Latin ones
run from $4$ to $D=3+d$; index $0$ is reserved for the time coordinate; hats denote background/unperturbed quantities. We also suppose that coordinates
$x^\mu,x^m$ have dimension of length. Therefore, the metric coefficients are dimensionless.

The background spacetime is perturbed by a matter source. In the weak field limit, the perturbed metric coefficients
\be{1.1} g_{MN}\approx \hat g_{MN}+h_{MN}\, ,\quad |h_{MN}|\ll 1\, . \ee
We consider a matter source consisting of two components. The first component is a compact body of mass $m$ and with mass density $\rho$. We assume that this
compact body corresponds to ordinary astrophysical objects. It is well known that the pressure inside these objects (e.g., inside our Sun) is much less than the
energy density. Therefore, we can neglect it. However, we do not know the pressure of these bodies in the internal space. So, we keep it and $\Omega$ is the
equation of state parameter in the internal space. We should note that the corresponding pressure is not connected with motion of gravitating masses, i.e.
$\Omega$ is the parameter of a body. The value of $\Omega$ should be restricted from observations. Since we are going to define the parameterized post-Newtonian
(PPN) parameter $\gamma$, it is sufficient in this case to calculate perturbations $h_{MN}$ (originated due to the gravitating mass) up to $O(1/c^2)$ \cite{EZ1}.
To do it, we need to keep the energy-momentum tensor (EMT) components up to $O(c^2)$ terms only, and for the nonzero components we get \cite{CEZ3}:
\ba{9}
&\textsf{T}_{00}&\approx\rho c^2, \quad \textsf{T}_{mn}\approx-\Omega \rho c^2\hg_{mn}\, ,\nn \\
&\textsf{T}&\approx \textsf{T}_{MN}\hg^{MN}=\rho c^2(1-\Omega d)\, .
\ea
Hereinafter, the energy-momentum tensor related to the gravitating body will be denoted by sans typeface.

The conservation equation for this tensor takes the form: $\nabla_{M}\textsf{T}^{M}_N=\hna_M\textsf{T}^{M}_N+o(c^2) = 0$ where $o(c^2)$ denotes terms which are
proportional to $c^n$ with $n<2$. We should drop such terms within the adopted accuracy. Then, the gravitating body is static\footnote{We consider only one
gravitating mass placed at the origin of a reference frame. That is, in such comoving frame, we disregard the spatial velocity of a body. Such simplification does
not affect the main results of our paper.}: $\hna_{M}\textsf{T}^{M0}=\de_0\textsf{T}^{0}_0=c^2\de_0\rho=0$ and is uniformly distributed/smeared over the internal
space: $\hna_M \textsf{T}^{M }_n=\hna_m \textsf{T}^{ m}_n=-\Omega c^2 \de_n\rho=0$ unless $\Omega=0$. In what follows, we will assume that the gravitating body is
localized at the origin in the external space and is smeared over the internal space: $\rho = m\delta({\bf r})/V_d$, where $V_d$ is the volume of the unperturbed
internal space and ${\bf r}=(x^1, x^2, x^3)$. Then, within the adopted accuracy, both of the matter source components satisfy  the conservation equations
separately.

The second material component is a static spherically symmetric scalar field $\phi$ with mass $\mu$ centered at the origin (see the footnote \ref{interaction}).
We assume that this scalar field is in its ground state in the internal space (there is no dependence on the extra spatial coordinates $x^m$). This field
satisfies the linearized Klein-Gordon equation $\de_r^2 \phi+(2/r)\de_r \phi-\mu^2\phi + O(h\phi)=0$ with the following solution\footnote{\label{interaction} To
provide a condition for the gravitating massive body and scalar field to stay at the same place (i.e. to be coupled), we can introduce an interacting term $\sim
\rho \phi$ into action. Since $\rho(\textbf{r}) \sim \delta(\textbf{r})$, this results in a delta function  in the rhs of the Klein-Gordon equation. Then, the
solution \rf{1.2} is valid in any point of the space, and the integration constant $\mathcal{C}$ is defined by a coupling constant. Moreover, the terms $\sim
\rho\phi$ will appear also in the EMT. They describe the energy of interaction of a massive body with scalar fields localized at other massive bodies.
As we can see from \rf{1.2}, such fields are exponentially suppressed for sufficiently large distances. Therefore, we will drop such interaction terms in the EMT.
In other words, we consider a one-body problem.}:
\be{1.2}
\phi(r)=\mathcal{C} e^{-\mu r}/r\, .
\ee
Here, the mass $\mu$ has dimension length $^{-1}$, $\mathcal{C}$ is some dimensional constant and $r\equiv\sqrt{(x^1)^2+(x^2)^2+(x^3)^2}$.

The EMT (which we denote with fraktur) of the scalar field reads:
\ba{2}
\mathfrak{T}_{MN}&=& \de_M\phi\de_N\phi-\cfrac{1}{2}\,\hg_{MN}\left[\de_L\phi\de^L\phi- \mu^2\phi^2\right]+O(h\phi^2)\, \nn,\\
&\phantom{}&\\
\label{3}
\mathfrak{T}&=&-\cfrac{d+2}{2}\,\de^L\phi\de_L\phi+\cfrac{1}{2}\,(d+4)\,\mu^2\phi^2+O(h\phi^2).
\ea
It can be easily realized that terms $O(h\phi^2)$ result in corrections of second order for the metric coefficients. Therefore, we should drop such terms within
adopted accuracy.

Taking into account the explicit form of $\phi$ \rf{1.2}, and introducing the spherical coordinates in the external space, we find that $\de_L\phi\de^L\phi=\hat
g^{rr}(\de_{r}\phi)^2=-\phi^2(\mu r+1)^2/r^2$, where $\hat g^{rr}=-1$. Then, the nonzero components of $\mathfrak{T}_{MN}$ take the form:
\ba{4}
\mathfrak{T}_{00}&\approx&\cfrac{\phi^2}{r^2} \left(\mu^2 r^2 + \mu r + \cfrac{1}{2}\,\,\right)\, ,\\
\label{5}\mathfrak{T}_{\bmu\bmu}&\approx&(\de_r\phi)^2(\de_{\bmu} r)^2-\mathfrak{T}_{00}, \quad (\de_{\bmu}r)^2=\cfrac{(x^{\bmu})^2}{r^2}\, ,\\
\label{6}
\mathfrak{T}_{\bmu\bnu}&\approx&(\de_{\bmu}\phi)(\de_{\bnu}\phi)=\cfrac{x^{\bmu}x^{\bnu}}{r^2}(\de_r\phi)^2\, , \quad \bmu\neq\bnu\, ,\\
\label{7}
\mathfrak{T}_{mn}&\approx&\hat g_{mn} \mathfrak{T}_{00}=\cfrac{\phi^2}{r^2} \left(\mu^2 r^2 + \mu r + \cfrac12\right)\hat g_{mn}.
\ea
Hereinafter, $\bar \mu, \bnu=1, 2, 3$ denote external space components.
The trace $\mathfrak{T}$, after substitution of the expression for $\de_L\phi\de^L\phi$ into (3), immediately takes the form
\be{8}
\mathfrak{T}\approx
\cfrac{\phi^2}{2 r^2} \left[(d+2) (\mu r + 1)^2 + (d+4) \mu^2 r^2\right]\, .
\ee

Both of matter components result in the spacetime perturbations \rf{1.1} where $h_{MN}$ can be found from
the linearized Einstein equation:
\be{10}
\delta R_{MN}=\kappa \left[T_{MN}-\cfrac{1}{d+2}\,T \hat g_{MN}\right]\equiv S_{MN},
\ee
where
\be{10.1}
T_{MN}
=\textsf{T}_{MN}+\mathfrak{T}_{MN}
\ee
is the total energy-momentum tensor and
\ba{11}
\delta R_{MN}&\approx& \cfrac{1}{2}\left[-\hat\nabla_L\hat\nabla^Lh_{MN}+\hat\nabla_L\hat\nabla_Mh^L_{N}\right.\nn \\
&+&\left.\hat\nabla_L\hat\nabla_Nh^L_{M}-\hat\nabla_N\hat\nabla_Mh^L_L\right]
\ea
is the perturbation of the Ricci tensor up to linear (with respect to $h_{MN}$) terms (see formulas (A.5)--(A.6) in \cite{CEZ1}). The prefactor $\kappa$ stands
for $2S_D G_{(D+1)}/c^4$, with $S_D$ being the total $D$-dimensional solid angle and $G_{(D+1)}$ being the constant of gravitational interaction in
$(D+1)$-dimensional spacetime. It is worth noting that the combination $\kappa \phi^2$ is dimensionless. Since equation \eqref{10} is linear with respect to
$h_{MN}$, the two material components $\textsf{T}_{MN}$ and $\mathfrak{T}_{MN}$ perturb the background geometry independently in the considered approximation.
Therefore, we can split $h_{MN}$ into the two contributions:
\be{11.1}
h_{MN}=\textsf{h}_{MN}+\mathfrak{h}_{MN}\, ,
\ee
where $\textsf{h}_{MN}$ is engendered by
\be{11.2}
\textsf{S}_{MN}\equiv \kappa\left[\textsf{T}_{MN}-\textsf{T}\hg_{MN}/(d+2)\right]
\ee
while $\mathfrak{h}_{MN}$ is produced by
\be{11.3}
\mathfrak{S}_{MN}=\kappa\left[\mathfrak{T}_{MN}-\mathfrak{T}\hg_{MN}/(d+2)\right]\, ,
\ee
respectively.

It  can be easily recognized that for the mass density $\rho = m\delta({\bf r})/V_d$ and the scalar field of the form \rf{1.2}, the combination $S_{MN}$ is static
($\de_0S_{MN}=0$), smeared over the internal space ($\hna_mS_{MN}=0$) as well as spherically symmetric in the external space ($S_{MN}({\bf r}, x^m)=S_{MN}(r,
x^m)$). Therefore, the corresponding metric perturbation $h_{MN}$ must possess the same properties, that is, $h_{MN}(x^M)=h_{MN}(r, x^m)$. Moreover, it is natural
to suppose that static and smeared matter sources preserve the block-diagonal structure of the spacetime:
\be{12}
h=\sum_{\mu=0}^3h_{\mu\mu}dx^\mu\otimes dx^\mu+h_{mn}dx^m\otimes dx^n.
\ee
Further, since $S_{\mu\nu} \, (\mu,\nu=0,1,2,3)$ does not depend on the internal coordinates, $h_{\mu\nu}$ also must not depend on $x^m$ ($\de_m h_{\mu\nu}=0, \,
m =4,\ldots , D=3+d$). Additionally, for the given matter sources, the EMT components $T_{mn}$ may depend on $x^m$ via $\hg_{mn}$ only. Therefore, we may suppose
that  $h_{mn}$ have the form $h_{mn}(r, x^m)=\xi(r)\hg_{mn}(x^m)$ (i. e. the metric on the internal space experiences a conformal perturbation). We will see below
that such prescriptions for the metric perturbations $h_{MN}$ are in full agreement with the linearized Einstein equations.

Similar to the splitting \rf{10.1}, we can distinguish in $h_{mn}(r, x^m)=\xi(r)\hg_{mn}$ contributions from different matter sources:
\be{12.1}
h_{mn} = \textsf{h}_{mn} + \mathfrak{h}_{mn} \equiv  \xi_f(r)\hg_{mn} + \xi_m(r)\hg_{mn}\, .
\ee
Obviously, $\xi_f+\xi_m=\xi$.

Taking all this into account, we can simplify \eqref{11} significantly:
\ba{13}
\delta R_{00}&\approx& \cfrac{1}{2}\,\triangle_3 h_{00}(r)\, ,\\
\delta R_{\bmu\bmu}&\approx& \cfrac{1}{2}\left[\triangle_3h_{{\bmu}{\bmu}}-2\de^2_{\bmu}h_{\bmu\bmu}\right. \nn \\
\label{14}&-&\left.\de_{\bmu}^2(h_{00}-\sum_{\bar\lambda}h_{\bar\lambda\bar\lambda}+\xi d )\right]\, ,\\
\delta R_{\bmu\bnu}&\approx&-\cfrac{1}{2}\left[\de_{\bmu}\de_{\bnu}h_{\bnu\bnu}+\de_{\bmu}\de_{\bnu}h_{\bmu\bmu}\right.\nn \\
\label{15}&+&\left.\de_{\bmu}\de_{\bnu}(h_{00}-\sum_{\bar\lambda}h_{\bar\lambda\bar\lambda}+\xi d)\right]\, ,\quad \bmu\neq\bnu\, ,\\
\label{16}\delta R_{mn}&\approx&\cfrac{1}{2}\,\hg_{mn}\triangle_3\xi\, ,
\ea
where $\triangle_3\equiv \sum_{\bmu=1}^3\de^2_{\bmu}$.

Now, let us calculate $S_{MN}=\textsf{S}_{MN}+\mathfrak{S}_{MN}$. Since $\textsf{S}_{MN}$ (defined by \rf{11.2}) is already known (see Eq. \rf{9}), we need to find $\mathfrak{S}_{MN}$ only:
\ba{17}
\mathfrak{S}_{00}&\equiv&\kappa\left[ \mathfrak{T}_{00}-\cfrac{1}{d+2}\,\mathfrak{T}\hg_{00}\right]=-\kappa\,\cfrac{\mu^2 \phi^2}{d+2}\, ,\\
\label{18}
\mathfrak{S}_{\bmu\bmu}&=&\kappa\left[\cfrac{(x^{\bmu})^2}{r^2}\,(\de_r\phi)^2+\cfrac{\mu^2 \phi^2}{d+2}\right]\, ,\\
\label{19}
\mathfrak{S}_{\bmu\bnu}&=& \kappa \left[\cfrac{x^{\bmu}x^{\bnu}}{r^2}\,(\de_r\phi)^2\right]\, ,\quad \bmu\neq\bnu\, ,\\
\label{20}
\mathfrak{S}_{mn}&=&\hg_{mn} \mathfrak{S}_{00}= -\kappa\cfrac{\mu^2 \phi^2}{d+2}\,\hg_{mn}\, .
\ea

\section{Metric coefficients}

\setcounter{equation}{0}

Combining \eqref{13}--\eqref{16} with \eqref{17}--\eqref{20}, we get the system of equations for $\mathfrak{h}_{MN}$
 (for the moment we set the constants $\mathcal{C}=\kappa=1$ to simplify the formulas):
\be{21}
\triangle_3 \mathfrak{h}_{00}=-\cfrac{2\mu^2 \phi^2}{d+2}\,, \quad \triangle_{3}\xi_f=-\cfrac{2\mu^2 \phi^2}{d+2}\, \quad\Rightarrow \quad \mathfrak{h}_{00}=\xi_f,
\ee
\ba{22} \triangle_3\mathfrak{h}_{{\bmu}{\bmu}}&-&2\de^2_{\bmu}\mathfrak{h}_{\bmu\bmu}-\de_{\bmu}^2(\mathfrak{h}_{00}-
\sum_{\bar\lambda}\mathfrak{h}_{\bar\lambda\bar\lambda}+\xi_fd ) \nn \\ &=&2\,\cfrac{(x^{\bmu})^2}{r^2}\,(\de_r\phi)^2+2\,\cfrac{\mu^2 \phi^2}{d+2}\, \ea
and
\ba{23}
&{}&\de_{\bmu}\de_{\bnu}\left[\mathfrak{h}_{\bnu\bnu}+\mathfrak{h}_{\bmu\bmu}
+(\mathfrak{h}_{00}-\sum_{\bar\lambda}\mathfrak{h}_{\bar\lambda\bar\lambda}+\xi_f d)\right]\nn \\
&=&-2\,\cfrac{x^{\bmu}x^{\bnu}}{r^2}\,(\de_r\phi)^2\, ,\quad \bmu\neq\bnu\, .
\ea

We denote the expression in square brackets in the lhs of \eqref{23} by $f_{\bmu\bnu}$. It can be easily seen that $f_{\bmu\bnu}$ is a function of $r$ only. Then
we can rewrite the lhs as follows:
\be{24}
\de_{\bmu}\de_{\bnu} f_{\bmu\bnu}(r)=\cfrac{x^{\bmu} x^{\bnu}} {r^2}\left[\de_r^2 f_{\bmu\bnu}-\cfrac{1}{r}\,\de_r f_{\bmu\bnu}\right], \quad \bmu\neq\bnu,
\ee
where we have used the evident relations of the form: $\de_{\bmu}f(r)= (df/dr)x^{\bmu}/r$. Hence, \eqref{23} results in \be{25} \de_r^2
f_{\bmu\bnu}-\cfrac{1}{r}\,\de_r f_{\bmu\bnu}=-2\,(\de_r\phi)^2, \quad \forall(\bmu\neq\bnu). \ee From \eqref{25} we conclude that all $f_{\bmu\bnu}$ are equal.
This is possible only provided that $\mathfrak{h}_{11}=\mathfrak{h}_{22}=\mathfrak{h}_{33}$, as it follows directly from the definition of $f_{\bmu\bnu}$.
Therefore, after summing over $\bmu$ in \eqref{22} and taking into account equation \rf{21} for $\mathfrak{h}_{00}$ we obtain:
\ba{29}
\triangle_3 \mathfrak{h}_{11}&=&\cfrac{1}{2}\,(\de_r\phi)^2-\cfrac{1}{2}\,\cfrac{d-2}{d+2}\,\mu^2 \phi^2\nonumber\\
&=& \cfrac{1}{2}\left[\cfrac{( \mu
   r+1)^2}{r^2}-\cfrac{d-2}{d+2}\,\mu^2 \right]\cfrac{e^{-2 \mu  r}}{r^2}.
\ea
The general solution of this equation has the form:
\ba{30}
&{}&\mathfrak{h}_{11}=C_1+\cfrac{C_2}{r}\nn \\
&-&\cfrac{d}{2(d+2)}\cfrac{\mu}{r}\,e^{-2\mu r}+\cfrac{e^{-2\mu r}}{4r^2}-\cfrac{d}{d+2}\,\mu^2\text{Ei}(-2\mu r),
\ea
where $\text{Ei}(x)$ is the exponential integral (see \cite{calc}, chapter 37). The natural boundary condition $\lim_{r\rightarrow\infty}\mathfrak{h}_{11}(r)=0$
implies $C_1=0$. Because there is no delta-like source in the rhs of \eqref{29}, $C_2$ is also zero.

Next, equations \eqref{21}, rewritten as
\be{31}
\triangle_3 \mathfrak{h}_{00}=\triangle_{3}\xi_f=-\cfrac{2\mu^2}{d+2}\, \cfrac{e^{-2\mu r}}{r^2},
\ee
under the same boundary conditions as for $\mathfrak{h}_{11}$, admit the solution:
\be{32}
\mathfrak{h}_{00}=\xi_f=-\cfrac{1}{d+2}\frac{\mu }{ r}\,e^{-2 \mu
   r}-\frac{2 }{d+2}\,\mu^2 \text{Ei}(-2
   \mu r)\, .
\ee To restore $\kappa\neq 1$ and $\mathcal{C}\neq 1$, we should multiply the right-hand sides of \eqref{30} and \eqref{32} by the combination $\kappa
\mathcal{C}^2$ which has the dimension of length$^{2}$. {It can be easily verified that these metric coefficients together with the expressions \rf{1.2} for $\phi$ satisfy equation \rf{25}.}
Taking into account Eq. \rf{12.1}, we obtain also the expression for $\mathfrak{h}_{mn}$:
\be{32.1}
\mathfrak{h}_{mn}=\xi_f\hg_{mn}\, .
\ee

The solutions for $\textsf{h}_{MN}$ were already obtained in \cite{CEZ1} (see formulas (11), (12)):
\be{33}
\textsf{h}_{00}=-\cfrac{2}{c^2}\,\cfrac{S_{D}G_{(D+1)}}{2\pi\hat V_d}\, \,\cfrac{1+(1+\Om)d}{d+2}\, \,\cfrac{m}{r}\,,
\ee

\be{34}
\textsf{h}_{\bmu\bnu}=\cfrac{1-\Om d}{1+(\Om+1)d}\,\textsf{h}_{00} \delta_{\bmu\bnu},
\ee

\be{35}
\xi_m=-\cfrac{2\Omega+1}{1+(\Om+1)d}\, \textsf{h}_{00}, \quad\Rightarrow\quad \textsf{h}_{mn}=\xi_m\hg_{mn}\, .
\ee

Finally, the total perturbations of metric coefficients  are given by formulas \rf{11.1} and \rf{12.1}.

\section{PPN parameter $\gamma$}

\setcounter{equation}{0}

It is well known that $h_{00}$ is related to the gravitational potential: $h_{00}=2\varphi/c^2$ \cite{Landau}. As it directly follows from \rf{33} and \rf{34},
the functions $\textsf{h}_{00}, \textsf{h}_{\bmu\bmu} \sim 1/r$ represent the pure Newtonian contributions to the corresponding metric coefficients $g_{00}$ and
$g_{\bmu\bmu}$, and the requirement that $\textsf{h}_{00}=2\varphi_N/c^2 = -(2/c^2)G_N m/r$ leads to a connection between the Newtonian $G_N$ and multidimensional
$G_{(D+1)}$ gravitational constants in the absence of the scalar field \cite{CEZ2}:
\be{37}
G_N=\cfrac{S_DG_{(D+1)}}{2\pi \hat V_d}\, \, \cfrac{1+(1+\Omega)d}{d+2}\, .
\ee
Additionally, if the scalar field is absent, the ratio
\be{35.1} \gamma=\frac{\textsf{h}_{\bmu\bmu}}{\textsf{h}_{00}} \, \ee
defines the parameterized post-Newtonian (PPN) parameter $\gamma$. In the case of Eqs. \rf{33} and \rf{34} we obtain
\be{35.2}
\gamma = \cfrac{1-\Om d}{1+(\Om+1)d}\, .
\ee
According to the experimental data of astronomical observations in the Solar system, $\gamma$ must be very close to $1$ \cite{Bertotti,Shapiro,Will2}. In General
Relativity, the considered ratio is exactly equal to unity: $\gamma=1$. For the expression \rf{35.2}, we can achieve this value only in the case of black
string/brane $\Omega =-1/2$ in agreement with \cite{CEZ1,EZ2,Braz}. That is, the gravitating masses should have tension in the internal space. However, we do not
know a physical reason for a gravitating body to have tension. The dust-like value $\Omega =0$ looks much more reasonable. Therefore, a natural question arises
whether the presence of a scalar field can provide such functions $\mathfrak{h}_{00}$ and $\mathfrak{h}_{\bmu\bmu}$ that they, first, contain contributions
demonstrating pure Newtonian behavior\footnote{In the PPN formalism, a static spherically symmetric line element in isotropic coordinates is parameterized in such
a way, that $h_{00}=-r_g/r$ and $h_{\bmu\bmu}=-\gamma r_g/r$, where $r_g=2G_Nm/c^2$ \cite{Will}. In the relativistic gravitational tests, $\gamma$ was estimated
for this parametrization. Therefore, to compare with astronomical limitations, we must find such a range of parameters where metric coefficients
$\mathfrak{h}_{00}$ and $\mathfrak{h}_{\bmu\bmu}$ behave as $1/r$.} at astrophysical scales and, second, contribute to the total ratio ${h}_{\bmu\bmu}/{h}_{00}$
in such a way that this ratio will be close to a unity for $\Omega=0$. Unfortunately, the answer is negative. To demonstrate it, we consider two limiting cases.

First, we consider the case of a ``heavy'' scalar field mass $\mu \gg r^{-1}$. Taking into account the asymptotic behavior of the exponential integral
$\operatorname{Ei}(x)\rightarrow \exp(x)/x$ when $|x|\rightarrow \infty$ (see \cite{calc}, chapter 37), we can conclude that in the limit $\mu r \gg 1$ the terms
$\mathfrak{h}_{00}$, $\mathfrak{h}_{11}$ decay much faster (i.e. exponentially) with distance than $\textsf{h}_{00}, \textsf{h}_{11}$. Then, at distances $r \gg
\mu^{-1}$ the metric coefficients $h_{00}\approx \textsf{h}_{00}$ and $h_{\bmu\bmu} \approx \textsf{h}_{\bmu\bmu}$, and for PPN parameter $\gamma$ we obtain
expression \rf{35.2} which requires the value $\Omega =-1/2$ to have the same accuracy as the General Relativity. The relativistic gravitational tests (e.g. the
deflection of light and the time delay of radar echoes) take place at distances $r \gtrsim R_{\odot}$ where $R_{\odot}$ is the radius of the Sun. Therefore, at
such distances scalar fields with masses $\mu \gtrsim 5.05\times 10^{-49}$g $\sim 2.83\times 10^{-16}$eV do not affect these gravitational tests.

Next, we consider the case of ultralight scalar field mass $\mu \ll r^{-1}$. Since the exponential integral asymptotically behaves as $\operatorname{Ei}(x)\approx
C_E+\ln(|x|)+x$, $|x|\ll1$ (here, $C_E$ stands for Euler's constant) \cite{calc}, then in the limit $\mu r\ll 1$ we get
\ba{38}
\mathfrak{h}_{00}&\approx& -\kappa\mathcal{C}^2\frac{\mu r}{(d+2)r^2}\left[1+2C_E\mu r -4 (\mu r)^2\right.\nn \\
&+& \left.2\mu r\ln (2\mu r)\right] \approx -\cfrac{\kappa\mathcal{C}^2}{d+2}\,\cfrac{\mu}{r}\, ,\\
\label{39}
\mathfrak{h}_{\bmu\bmu} &\approx& \kappa\mathcal{C}^2\frac{1}{r^2}\left(\frac{1}{4} -\frac{2}{2(d+2)}\mu r\right) \approx  \frac{\kappa\mathcal{C}^2}{4r^2}\, ,
\ea
where we have restored the dimensional prefactor $\kappa\mathcal{C}^2$. For the total metric coefficients we obtain:
\be{38.1} h_{00}=\textsf{h}_{00}+\mathfrak{h}_{00}\approx -\cfrac{\kappa c^2}{2\pi \hat V_d}\cfrac{1+(1+\Omega)d}{d+2}\,\cfrac{m}{r}-\cfrac{\kappa
\mathcal{C}^2}{d+2}\,\cfrac{\mu}{r}. \ee
and
\be{39.1}
h_{\bmu\bmu}=\textsf{h}_{\bmu\bmu}+\mathfrak{h}_{\bmu\bmu}
\approx-\cfrac{\kappa c^2}{2\pi \hat V_d}\,\cfrac{1-\Omega d}{d+2}\,\cfrac{m}{r}+\cfrac{\kappa\mathcal{C}^2}{4 r^2}\, .
\ee

As we mentioned above, metric coefficients should demonstrate the Newtonian $1/r$ behavior. This means that in the expression \rf{39.1} for $h_{\bmu\bmu}$ the
second term with $1/r^2$ should be much less than the first one with $1/r$. Then, we get the additional condition for $r$:
\be{43}
r\gg \frac{\pi(d+2)}{2}\, \cfrac{\mathcal{C}^2\hat V_d}{mc^2(1-\Omega d)}\, .
\ee
Therefore, $h_{\bmu\bmu}\approx \textsf{h}_{\bmu\bmu}$.

To be in agreement with the relativistic gravitational tests (i.e. to have $\gamma \approx 1$), we must equate $h_{00}$ and $h_{\bmu\bmu}$. This results in the
relation:
\be{46}
-d\left(1+2\Omega\right)mc^2=2\pi \hat V_d\mathcal{C}^2 \mu\, .
\ee
This equation shows that the mass of scalar field $\mu$ is defined by the mass of the gravitating body. Since the scalar field and the gravitating body are
independent, such condition looks very artificial. Moreover, the condition of positivity\footnote{It is worth noting that in the exotic case of negative mass
$\mu<0$, the value $\Omega =0$ is possible.} of $\mu$: $\Omega<-1/2$ excludes the sought value $\Omega =0$. Therefore, we can satisfy \rf{46} only if both the lhs
and the rhs are equal to zero simultaneously. Obviously, this is possible either when $\mu=0$ (massless field), or when $\mathcal C=0$ (absence of the field at
all), but in both cases $\Omega=-1/2$. It follows from above that a massless scalar field does not affect the gravitational test for the PPN parameter $\gamma$ in
the Solar system subject to the condition \rf{43}, where $r$ is replaced by $R_{\odot}$ and $m$ by $M_{\odot}$.

Therefore, the presence of the scalar field (massive or massless) does not improve the situation with respect to the parameter $\Omega$. In all cases, we arrive
at the same conclusion that to be in agreement with the relativistic gravitational tests the gravitating mass should have tension: $\Omega=-1/2$.

\section{Conclusion}

\setcounter{equation}{0}

In this paper, we have considered the effect of a scalar field on the PPN parameter $\gamma$ in a multidimensional Kaluza-Klein model with Ricci-flat internal
space. In this model, the gravitational field is created by two coupled sources. First, this is a point-like massive body which has a dust-like equation of state
in the external space and an arbitrary parameter $\Omega$ of the equation of state in the internal space. In the static limit and within the adopted accuracy, the
massive gravitating body is smeared over the internal space. The second source is a static spherically symmetric massive scalar field centered at the origin where
the point-like massive body is situated. We have assumed also that this scalar field is in its ground state in the internal space and does not depend on the extra
spatial coordinates. In the linear approximation, we have calculated the perturbed metric coefficients for this model. We have used these perturbations to
calculate the parameterized post-Newtonian parameter $\gamma$. This PPN parameter is well constrained by the gravitational tests (e.g. the deflection of light and
the time delay of radar echoes) in the Solar system. According to these tests, $\gamma$ is extremely close to unity, which is in very good agreement with the
General Relativity prediction \cite{Bertotti,Shapiro,Will2}. In Kaluza-Klein models with Ricci-flat internal space, the same value of $\gamma$ as in the General
Relativity (i.e. $\gamma =1$), can be achieve only for black brane value $\Omega=-1/2$ \cite{EZ1,CEZ1,CEZ2}. However, we consider this value as not very realistic
since there is no any physically reasonable motivation for it up to now. Therefore, in the present paper we have investigated a possibility with the help of
scalar field to shift $\Omega$ to a more natural value $\Omega=0$. Simultaneously, we have tried to keep the PPN parameter $\gamma$ close to unity.  We have shown
that we can fulfill the latter condition for massive as well as massless{\footnote{It is well known that massless fields can result in a potentially dangerous situation if they are coupled to matter. This usually leads to the fifth-force problem. We can see it e.g. from Eq. \rf{38.1} obtained in the case of light mass scalar field. Here, we get the additional contribution of the order of 1/r to the gravitational potential. With respect to the PPN parameter $\gamma$, we eliminated this danger by putting $\mu =0$ and $\Omega=-1/2$. However, such fifth force can manifest itself in other experiments. This can result in additional restrictions on the constant of interaction $ \mathcal{C}$.}} scalar field. For example, in the Solar system to have $\gamma \approx 1$, the mass of
scalar field should be $\mu \gtrsim 5.05\times 10^{-49}$g $\sim 2.83\times 10^{-16}$eV. Unfortunately, in all cases, we have arrived at the same conclusion that
to be in agreement with the relativistic gravitational tests, the gravitating mass should have tension: $\Omega=-1/2$.


\section*{Acknowledgements}

We would like to thank Maxim Eingorn for his contribution during the initial phase of work as well as for stimulating discussions and valuable comments.


\end{document}